\documentclass[12pt,preprint]{article}
\usepackage{amssymb}
\usepackage{amsmath}
\usepackage[dvips]{graphicx}

\renewcommand{\vec}[1]{{\bf #1}}
\newcommand{\eqb}{\begin{equation}}
\newcommand{\eqe}{\end{equation}}
\newcommand{\dmb}{\begin{displaymath}}
\newcommand{\dme}{\end{displaymath}}

\newcommand{\eab}{\begin{eqnarray}}
\newcommand{\eae}{\end{eqnarray}}

\newcommand{\be}{\begin{equation}}
\newcommand{\ee}{\end{equation}}

\begin{document}

\begin{titlepage}
\begin{flushright}
KA-TP-17-2008 
\end{flushright}
\vspace{0.6cm}

\begin{center}
\Large{Center-vortex loops with one selfintersection}

\vspace{1.5cm}

\large{Julian Moosmann and Ralf Hofmann}
\end{center}
\vspace{1.5cm} 

\begin{center}
{\em Institut f\"ur Theoretische Physik\\ 
Universit\"at Karlsruhe (TH)\\ 
Kaiserstr. 12\\ 
76131 Karlsruhe, Germany}
\end{center}
\vspace{1.5cm}

\begin{abstract}
We investigate the 2D behavior of one-fold selfintersecting, 
topologically stabilized center-vortex loops in the confining phase of
an SU(2) Yang-Mills theory. This coarse-graining is described by
curve-shrinking evolution of center-vortex loops immersed in a flat 
2D plane driving the renormalization-group flow of an effective
`action'. We observe that the system evolves into a highly ordered 
state at finite noise level, and we speculate that this feature is connected
with 2D planar high $T_c$ superconductivity in
$FeAs$ systems.  

\end{abstract} 

\end{titlepage}

\section{Introduction}

The idea of a nontrivial ground state being responsible 
for the emergence of `elementary' particles is a rather old 
one: Already Lord Kelvin proposed 
that atoms and molecules should be considered 
knotted lines of vortices representing distortions in 
a universal medium (or ground state) -- the ether \cite{Kelvin}. As we know now, 
the physics of atoms and molecules is described in terms of a much more efficient and elegant framework: Quantum
mechanics. The agent responsible for the 
chemical bond -- Kelvin's electron -- is considered a spinning point 
particle in quantum mechanics, and this yields an 
excellent description of atomic physics, 
collider physics, and for the bulk of situations in 
condensed matter physics.  

There are, however, theoretical discrepancies with the 
concept of the electron being a point particle, and there are 
exceptional experimental situations pointing to the limitations of this 
concept to describe reality. As for the former, we have the old problem of a diverging classical 
selfenergy not resolved in quantum 
electrodynamics where the electron mass is introduced 
as a free parameter whose running with 
resolution needs an experimental boundary condition. On the other 
hand, the two-dimensional dynamics of strongly 
correlated electrons in condensed matter physics signals  
the relevance of nonlocal effects possibly related to the 
nontrivial anatomy of the electron 
becoming relevant in collective phenomena 
\cite{Muller1986,Anderson2005}. Also, recent 
high-temperature plasma experiments indicate 
unexpected explosive behavior not unlikely 
related to the mechanism for lepton emergence, see \cite{Zpinch} and 
references therein. 

Recent developments in understanding the 
confining phase of an SU(2) Yang-Mills 
theory suggest that Kelvin's 
ideas may actually be realized in 
Nature, see also \cite{FaddeevNiemi}. The authors of 
\cite{FaddeevNiemi} construct a plausible effective low-energy action for the 
4D SU(2) Yang-Mills theory with solutions to the associated field
equations representing closed confining strings 
knotted into stable solitons. In the thermodynamic approach of 
\cite{Hofmann2005} the emergence 
of magnetic center-vortex loops (CVLs) is related
to discontinuous phase changes 
of a complex order parameter for confinement across the (downward) Hagedorn 
transition and the fact that no magnetic 
charges exist where these flux lines could end. Also, it was discussed in
\cite{Hofmann2005} how the 
locations of topologically stabilized selfintersection represent 
isolated, spinning magnetic charges \footnote{Notice that with respect to
the electromagnetic U(1) of the Standard Model there is a dual
interpretation of magnetic charges emerging in an SU(2) Yang-Mills theory.}. 

In our previous article \cite{MH2008} we have 
investigated 
the sector with $N=0$ selfintersections by 
considering a resolution dependent 
ensemble average. The corresponding weight-functional 
is defined purely in terms of the planar curves's 
geometry. The resolution dependence of this geometry, 
in turn, is determined by a 
curve-shrinking equation (heat-equation)
\cite{GageHamilton,Grayson}. The validity of this description of spatial 
coarse-graining is motivated from considerations
relating local curvature with the direction and speed of `motion' of the associated 
line-segment. The requirement that 
the partition function over a given ensemble of 
planar curves is invariant under a change of 
the resolution then yields the 
renormalization-group evolution 
of the weight-functional which is written as the exponential 
of an `action'. Here the term `action' is slightly misleading 
since we do not aim at describing the time-evolution of the 
system by demanding stationarity of the `action' under curve variation. 
To do the latter, a model, which 
relates resolution and time (being a macroscopic concept associated with
the measuring apparatus), needs to be introduced. We thus 
regard resolution over time as the more fundamental quantity to describe
certain subatomic systems. Our observation 
is that the effective `action' exhibits a transition 
towards dilational invariance after a 
finite, critical decrease of resolution. 
On average, CVLs with $N=0$ are shrunk to circular 
points for a resolution less than the critical value which defacto removes them 
from the spectrum and thus generates an asymptotic mass gap\footnote{CVLs with $N>0$ are massive \cite{Hofmann2005,Hofmann2007}.}. Knowing
the evolution of the weight-functional, one is in a position to compute
the resolution dependence of `observables' as ensemble averages of the
associated (nonlocal or local) `operators'. As for the evolution 
of the initially sharp center-of-mass position, 
we observe a spread of the variance with decreasing resolution
saturating at a finite value. This is 
similar to the unitary free-particle evolution of a position eigenstate
in quantum mechanics.      

\begin{figure}
\centering
\vspace{3cm}
\includegraphics{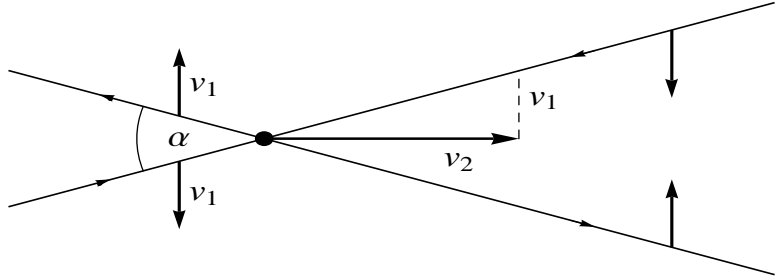}
\caption{\protect{\label{Fig-1}}Points on the center flux lines moving
  oppositely on a line perpendicular to the bisecting line of the angle 
$\alpha$ with velocity modulus $v_1$. For sufficiently small $\alpha$ 
the velocity modulus $v_2$ of the intersection point is superluminal: 
$v_2=v_1\cot{\frac{\alpha}{2}}$.}
\end{figure}

The purpose of the present article is to extend the procedure 
of \cite{MH2008} to the case of $N=1$. We now have a 
singled-out point on the curve: the location of 
the selfintersection where practically the entire mass of the soliton
resides \cite{Hofmann2005}. Setting the Yang-Mills 
scale $\Lambda$ of the SU(2) theory equal to the electron mass $m_e=511\,$keV,
which in turn determines the mass of the intersection point, 
we interprete this soliton as an electron or a positron 
\cite{Zpinch,Hofmann2007}. In the presence of a static 
electric or magnetic background field it is physically possible to lift
the two-fold degeneracy w.r.t. the two possible directions of
center-flux: The soliton exhibits a two-fold spin degeneracy. 
Notice that as long 
as both wings of center flux are of finite size the 
position of the intersection point can be shifted 
at almost no cost of energy. In particular, 
if the inner 
angle $\alpha$ between in- and out-going 
center-flux at the intersection is sufficiently small then a motion of points on the
vortex line directed perpendicular to the bisecting line of the angle
$\alpha$ easily generates 
a velocity of the intersection point which exceeds the speed of light, see
Fig.\,\ref{Fig-1}. Recall, that the path-integral formulation of quantum
mechanics admits such superluminal motion in the sense that the
according trajectories sizably contribute to transition amplitudes. 

The paper is organized as follows. In Sec.\,\ref{CFW} we discuss the
physics associated with the emergence of topologically stabilized
CVLs with intersection number $N=1$, and how their
spatial 2D coarse-graining is captured by a curve-shrinking flow. Some
mathematical results on the properties of this flow for 
immersed curves, which are relevant for our subsequent numerical
analysis, are briefly discussed. Also, we repeat our discussion in
\cite{MH2008} of how the renormalization-group flow of an effective
`action' is driven by the curve-shrinking evolution of the members of a
given ensemble of curves. In Sec.\,\ref{ROF} we explain our numerical 
analysis concerning the computation of the effective `action', the
variance of the location of the selfintersection, and the entropy
associated with a given ensemble. Finally, in Sec.\,\ref{SIC} we
summarize our results and interprete them in view of certain 2D layered, 
quasimetallic systems exhibiting high-$T_c$ superconductivity.     

\section{Conceptual framework\label{CFW}}

\subsection{Selfintersecting center-vortex loops}
\label{SCVL}

\begin{figure}
\centering
\vspace{9cm}
\includegraphics{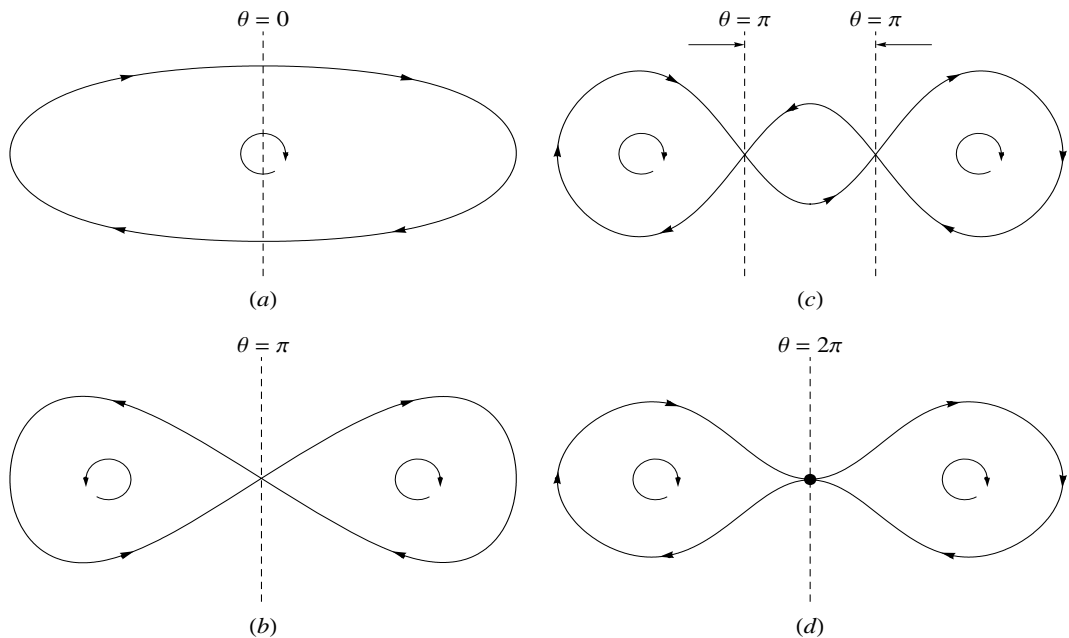}
\caption{\protect{\label{Fig-2}}(Topological) transition from the $N=0$
  sector (a), (b), (c) to the $N=1$ sector (d) by twisting and
  subsequent capture of a magnetic (anti)monopole in the core of 
  the final intersection. Arrows indicate the direction of center flux.}
\end{figure}

The transition from the non-selfintersecting to the selfintersecting CVL sector 
is by twisting of non-selfinteresecting curves. The emergence of a localized 
(anti)monopole in the process is due to its capture by oppositely
directed center fluxes in the intersection core (eye of the storm). 
By a rotation of the left half-plane in Fig.\,\ref{Fig-2}(a) by an angle of
$\pi$, see Fig.\,\ref{Fig-2}(b), each wing of the CVLs forms 
a closed flux loop by itself thereby introducing equally 
directed center fluxes at the intersection point. This does 
not allow for an isolation of a single, spinning (anti)monopole in the core of
the intersection and thus is topologically equivalent to the 
untwisted case Fig.\,\ref{Fig-2}(a). However, another rotation of the
left-most half-plane in Fig.\,\ref{Fig-2}(c) introduces an intermediate 
loop which by shrinking is capable of isolating a spinning
(anti)monopole due to oppositely directed center fluxes. Notice that in
the last stage of such a shrinking process (short distances 
between the cores of the flux lines), where propagating dual gauge 
modes are available\footnote{On large distances these modes are
  infinitely massive which is characteristic of the confining phase.},
there is repulsion due to Biot-Savart which needs to be overcome. This
necessitates an investment of energy manifesting itself in terms of the
mass of the isolated (anti)monopole (eye of the storm). Alternatively, 
the emergence of an isolated (anti)monopole is possible by a simple pinching 
of the untwisted curve, again having to overcome local repulsion in the
final stage of this process. 

For the analysis performed in the present work we solely 
regard the situation depicted in Fig.\,\ref{Fig-2}(d) and thus no longer
need to discuss the direction of center flux within a given curve
segment: This is not relevant for the process of a spatial
coarse-graining microscopically described by the same curve-shrinking
flow as applied to sector with $N=0$ \cite{MH2008}. 

\subsection{Euclidean curve shrinking flow}
\label{ECSF}

\begin{figure}
\begin{center}
\vspace{6cm}
\includegraphics{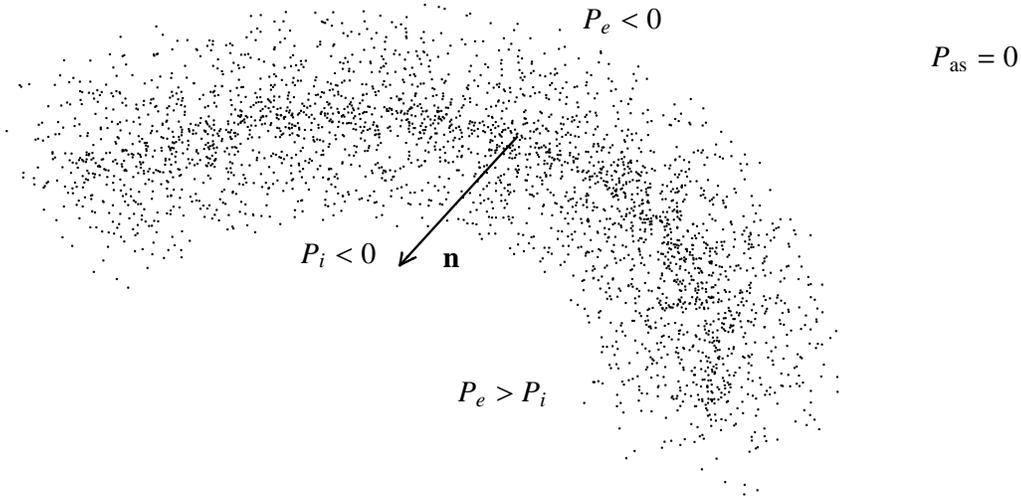}
\end{center}
\caption{\protect{\label{Fig-3}}Highly space-resolved snapshot of a CVL
  segment. The pressure $P_i$ in the region pointed to by the normal
  vector $\vec{n}$ is more negative than the pressure $P_e$ thus leading to a motion of the segment along $\vec{n}$.}
\end{figure}

Notice that by immersing an SU(2) CVL with finite core size $d$ and mass $m_d$ of the dual gauge field into a 
flat 2D surface at $m_D<\infty$, $d>0$, a hypothetic observer measuring a positive (negative) 
curvature of a segment of the vortex line experiences more (less)
negative pressure in the intermediate vicinity of this curve segment
leading to its motion towards (away from) the observer, see Fig.\,\ref{Fig-3}. 
The (inward directed) speed of a point in the core of the vortex will be a monotonic function of the 
curvature at this point. On average, this shrinks the CVL. Alternatively, one may {\sl globally}
consider the limit $m_D\to\infty$, $d\to 0$,
that is, the confining phase of an SU(2) Yang-Mills theory, 
but now take into account the effects of an environment which 
{\sl locally} relaxes this limit (by collisions) and thus also induces
curve shrinking. This situation is described by the 
following equation for a flow in the (dimensionless) parameter 
$\tau$    
\eqb
\label{CSt}
\partial_\tau \vec{x}=\frac{1}{\sigma}\,\partial^2_s \vec{x}\,,
\eqe
where $s$ is arc length, $\vec{x}$ is a point on the CVL in the plane, and 
$\sigma$ is a string tension effectively 
expressing the distortions induced by the environment. After a rescaling,
$\hat{x}\equiv\sqrt{\sigma}\vec{x}$, $\xi=\sqrt{\sigma}s$,
Eq.\,(\ref{CSt}) assumes the following form
\eqb
\label{CSta}
\partial_\tau \hat{x}(u,\tau)=\partial^2_\xi \hat{x}=k(u,\tau)\vec{n}(u,\tau)\,,
\eqe
where $u$ is a (dimensionless) curve parameter, $\vec{n}$ 
the (inward-pointing) Euclidean unit normal, $k$ the scalar curvature, defined as 
\eqb
\label{curvdef}
k\equiv\left|\partial^2_\xi \hat{x}\right|
=\left|\frac{1}{|\partial_u\hat{x}|}\partial_u\left(\frac{1}{|\partial_u\hat{x}|}\partial_u\hat{x}\right)\right|\,,
\eqe
$|\vec{v}|\equiv\sqrt{\vec{v}\cdot\vec{v}}$, and $\vec{v}\cdot\vec{w}$ 
denotes the Euclidean scalar product of the vectors $\vec{v}$ and
$\vec{w}$. In the following we resort to a slight abuse of notation by
using the same symbol $\hat{x}$ for the functional dependence on $u$ or
$\xi$. 

We now consider curves with one selfintersection, that is $N=1$, in the sense of the
stable situation of Fig.\,\ref{Fig-2}(d). This situation was
mathematically analysed in \cite{GraysonII}. Since the direction of center
flux is inessential for the shrinking process we may actually treat this
situation in a way as depicted in Fig.\,\ref{Fig-2}(b), where the curve is 
defined to be a smooth immersion into the plane with exactly
one double point and a total rotation number zero, $\int_{0}^{L} k
\, d\xi = 0$. Here the (dimensionless) curve length $L$ is given by the smooth
integration $L(\tau)=\int_0^{L(\tau)} d\xi=\int_0^{2\pi}
du\,\left|\partial_u\hat{x}(u,\tau)\right|$. Notice that 
this is topologically distinct from the case 
Fig.\,\ref{Fig-2}(d) where one encounters a nonvanishing 
rotation number which is not smoothly deformable to zero. 

In the $N=0$ case a smooth, embedded curve 
shrinks to a circular point under the flow for $\tau \nearrow
T<\infty$ \cite{GageHamilton,Grayson}. That is, the isoperimetric ratio approaches $4\pi$ from
above. The curve in situation Fig.\,\ref{Fig-2}(b) 
separates the plane into three disjoint areas two of
which are finite and denoted by $A_1$ and $A_2$. We understand by $T$ the
finite, critical value of $\tau$ where either $A_1$ or $A_2$ or both 
vanish. This corresponds to a singularity encountered and thus
terminates the flow.

Recall that in the $N=0$ case the rate of area change is a constant,
$\frac{dA}{d\tau}=-2\pi$. This is no longer true for $N=1$. 
However, we have that
\eqb
\label{diffA}
A_1(\tau)-A_2(\tau)=\mbox{const}\,. 
\eqe
Also, for $N=1$ we have in comparison to the $N=0$ case the more relaxed constraint that
$-4\pi\le\frac{d(A_1+A_2)}{d\tau}\le -2\pi$.

In contrast to the $N=0$ case the isoperimetric ratio for the $N=1$ case
is bounded for $\tau \nearrow T$ if and only if $A_1\not=A_2$. Notice that the case
$A_1=A_2$ physically is extremely fine-tuned.   

\subsection{Effective `action'}
\label{EA}

We now wish to interprete curve-shrinking as a Wilsonian
renormalization-group flow taking place in the $N=1$ 
CVL sector in the sense defined in Sec.\,\ref{ECSF}. A partition function, 
defined as a statistical average (according to a suitably defined weight) over $N=1$ 
CVLs, is to be left invariant under a decrease of the 
resolution determined by the flow parameter $\tau$. 
Notice that, physically, $\tau$ is interpreted as a strictly
monotonic decreasing (dimensionless) 
function of a ratio $\frac{Q}{Q_0}$ where $Q$ ($Q_0$) are mass 
scales associated with an actual (initial) resolution applied to the
system. The role of $Q$ can also be played by the 
finite temperature of a reservoir coupled to the system. 

To device a geometric ansatz for the 
effective `action' $S=S[\hat{x}(\tau)]$, which is a functional of the
curve $\hat{x}$ representable in terms 
of integrals over local densities in $\xi$ (reparametrization
invariance), the following reflection on symmetries is in order.  
(i) scaling symmetry $\hat{x}\to \lambda\hat{x}\,,\ \ \lambda\in{\mathbf
  R}_+$: For $\lambda\to\infty$, implying
$\lambda L\to\infty$ at fixed $L$, the `action' $S$ should 
be invariant under further finite rescalings (decoupling of the fixed
length scales $\sigma^{-1/2}$ and $\Lambda^{-1}$), 
(ii) Euclidean point symmetry of the plane (rotations, translations
and reflections about a given axis): Sufficient but not necessary for
this is a representation of $S$ in terms of integrals over 
scalar densities w.r.t. these symmetries. That is, 
the `action' density should be expressible as a
series involving products of Euclidean scalar products of $\frac{\partial^n}{\partial
    \xi^n}\hat{x}\,,\ \ n\in\mathbf{N}_+\,,$ or constancy. 
However, an exceptional scalar integral over a nonscalar density can be 
deviced. Consider the area $A$, calculated as 
\eqb
\label{area}
A=\left|\frac12\,\int_0^{2\pi} d\xi\,\hat{x}\cdot\vec{n}\right|\,.
\eqe
The density $\hat{x}\cdot\vec{n}$ in Eq.\,(\ref{area}) is not a scalar under
translations. 

We now resort to a factorization ansatz as 
\eqb
\label{effectactdef}
S=F_c\times F_{nc}\,,
\eqe
where in addition to Euclidean point symmetry $F_c$ ($F_{nc}$) is (is not) invariant 
under $\hat{x}\to \lambda\hat{x}$. In principle, 
infinitely many operators can be defined to contribute to $F_c$. 
Since the evolution homogenizes the curvature except 
for a small vicinity of the intersection point higher derivatives of 
$k$ w.r.t. $\xi$ should not be of importance. We expect this to be true 
also for Euclidean scalar products involving higher 
derivatives $\frac{\partial^n}{\partial \xi^n}\hat{x}$. 
To yield conformally invariant expressions such integrals need to be 
multiplied by powers of $\sqrt{A}$ and/or $L$ or the inverse of 
integrals involving lower derivatives. At this stage, we are 
not capable of constraining the expansion in derivatives by additional physical or
mathematical arguments. To be pragmatic, 
we simply set $F_c$ equal to the isoperimetric ratio:
\eqb
\label{explFc}
F_c(\tau)\equiv\frac{L(\tau)^2}{A(\tau)}\,.
\eqe
We conceive the nonconformal factor $F_{nc}$ in $S$ as a formal Taylor expansion in inverse powers of 
$L$ or $A\equiv A_1+A_2$ due to the property of conformal invariance for $L,A\to\infty$.

Since we regard the renormalization-group evolution of the
effective `action' as induced by the flow of an ensemble of curves, where the evolution of each member is
dictated by Eq.\,(\ref{CSta}), we allow for an explicit $\tau$
dependence of the coefficient $c$ of the lowest nontrivial power
$\frac{1}{L}$ or $1/A$. In principle, this sums up the contribution to $F_{nc}$ of certain 
higher-power operators which do not exhibit an explicit $\tau$
dependence. Hence we make the following ansatz 
\eqb
\label{explFnc}
F_{nc}(\tau)=1+\frac{c(\tau)}{L(\tau)}\,.
\eqe
The initial value $c(\tau=0)$ is determined from a physical 
boundary condition such as the mean length $\bar{L}$ at $\tau=0$.

\subsection{Geometric partition function}
\label{GPF}

Let us now numerically investigate the effective `action'
$S[\hat{x}(\tau)]$ resulting from a 
partition function $Z$ w.r.t. a nontrivial ensemble $E$. The latter is defined as the average
\eqb
\label{PartZM}
Z=\sum_{i} \exp\left(-S[\hat{x}_i(\tau)]\right)
\eqe
over the ensemble $E=\{\hat{x}_1,\cdots\}$. 
Let us denote by $E_M$ an ensemble consisting of $M$ curves where 
$E_M$ is obtained from $E_{M-1}$ by adding a new curve
$\hat{x}_M(u,\tau)$. The effective `action' $S$ in
Eq.\,(\ref{effectactdef}) (when associated with the ensemble $E_M$ we
will denote it by $S_M$) is determined by the function $c_M(\tau)$,
compare with Eq.\,(\ref{explFnc}), whose 
flow follows from the requirement of $\tau$-independence of $Z_M$:
\eqb
\label{renflow}
\frac{d}{d\tau}Z_M=0\,.
\eqe 
This is an implicit, first-order ordinary differential equation 
for $c(\tau)$ which needs to be supplemented with an initial condition
$c_{0,M}=c_M(\tau=0)$. A natural initial condition is to 
demand that the quantity 
\eqb
\label{barL}
\bar{L}_M(\tau=0)\equiv\frac{1}{Z_M(\tau=0)}\sum_{i=1}^M
L[\hat{x}_i(\tau=0)]\,\exp\left(-S_M[\hat{x}_i(\tau=0)]\right)
\eqe
coincides with the geometric mean $\tilde{L}_M(\tau=0)$ defined as 
\eqb
\label{geommean}
\tilde{L}_M(\tau=0)\equiv\frac{1}{M}\sum_{i=1}^M L[\hat{x}_i(\tau=0)]\,.
\eqe
From $\bar{L}_M(\tau=0)=\tilde{L}_M(\tau=0)$ a value for $c_{0,M}$
follows. We also have considered a modified factor
\eqb
\label{explFncmod}
F_{nc}(\tau)=1+\frac{c(\tau)}{A(\tau)}\,.
\eqe
in Eq.\,(\ref{effectactdef}). While the ansatz for the geometric effective `action' 
thus is profoundly different for such a modification 
of $F_{nc}(\tau)$ physical results such 
as the evolution of the variance of the intersection agree
remarkably well, see Sec.\,\ref{ROF}.

\section{Results of simulation}
\label{ROF}

\subsection{Preparation of ensembles}

Similar as in \cite{MH2008} we normalize all curves to have the same
initial area $A_0=A_{0,1}+A_{0,2}$ and, since we are now interested in the position of the
intersection where the (anti)monopole is localized, we have applied a
translation to each curves in the ensembles $E_M$ such that the location
of the intersections initially coincide with the origin. 

Since the critical value $T$ of the flow parameter $\tau$ 
varies from curve to curve we order the members of the maximal-size ensemble 
$E_{M=16}$ into subensembles $E_{M<16}$ such that $T_{i=1}\ge T_{i=2}\ge\cdots\ge
T_M$. The types of ensembles $E_M$ obtained in this way are referred to as
$T$-ordered. We also have performed all simulations with ensembles
$E^\prime_{M<16}$ whose members are picked randomly from $E_{M=16}$ and
have obtained strikingly similar results for ensemble averages of
`observables' using $E_{M<16}$ and
$E^\prime_{M<16}$ for the $\tau$ evolution to the left of
$\tau=\min\{T_i|\hat{x}_i\in E^\prime_{M<16}\}$. 

\begin{figure}
\centering
\vspace{9.5cm}
\includegraphics{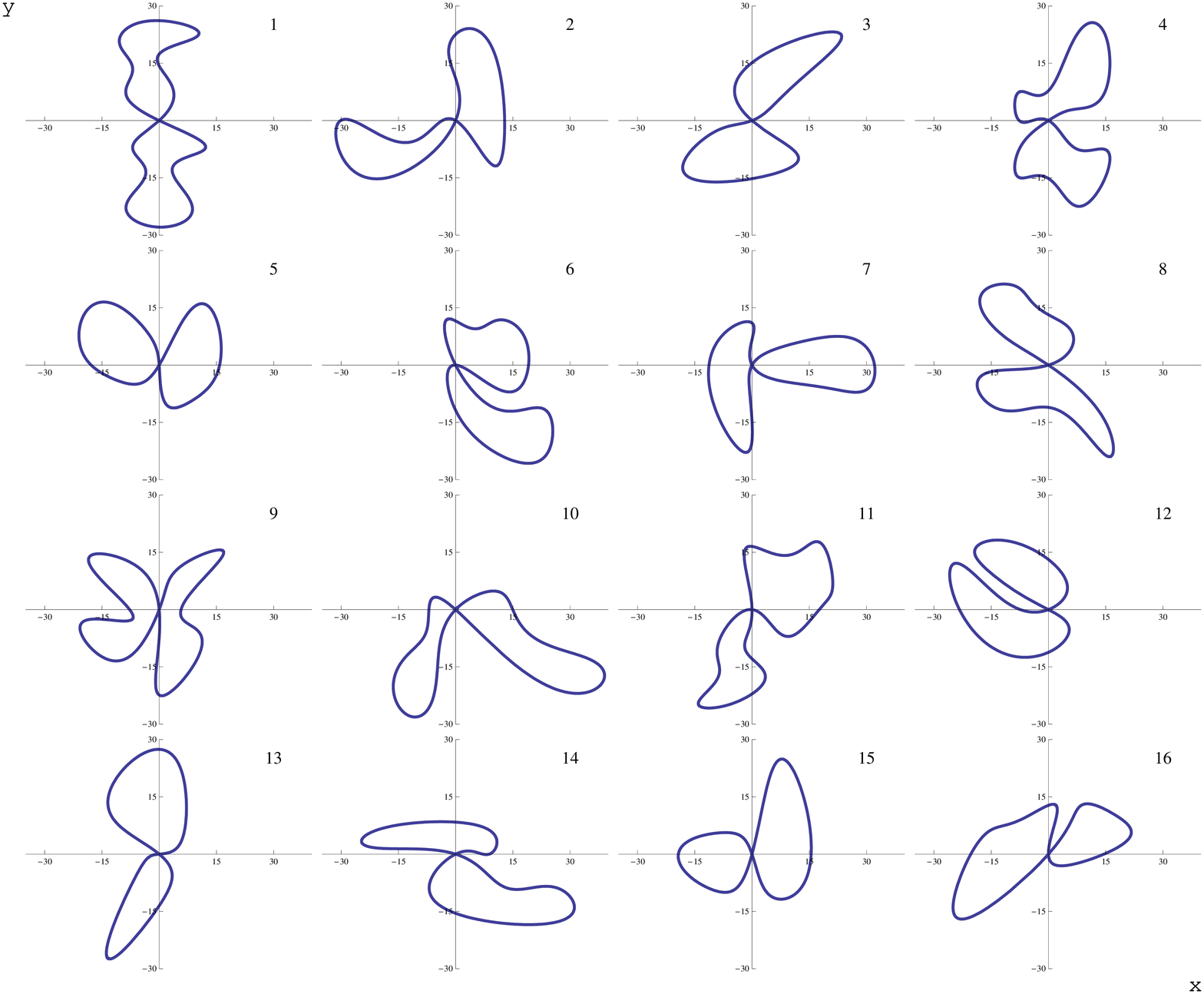}
\caption{\protect{\label{Fig-4}}Initial curves $\hat{x}_i(u,\tau=0)$ contributing to the
  ensemble $E_{M=16}$. The intersection points coincide with
  the origin, and all curves have the same area $200\,\pi$. By
  definition $E_{M=16}$ is $T$-ordered.}
\end{figure}

\begin{figure}
\centering
\vspace{8cm}
\includegraphics{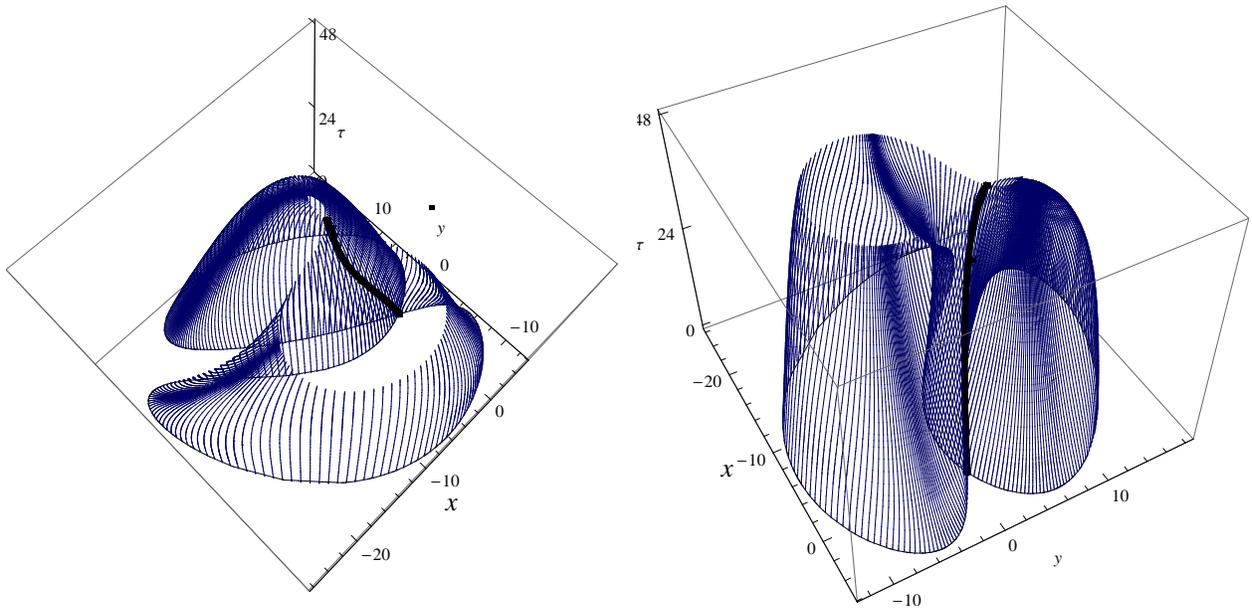}
\caption{\protect{\label{Fig-5}}Plot of the evolution of an $N=1$ CVL (curve 12 of Fig.\,\ref{Fig-4}) under 
Eq.\,(\ref{CSta}). The thick central line indicates the trajectory 
  of the intersection point which coincides with the origin at $\tau=0$.}
\end{figure}

\begin{figure}
\centering
\vspace{7.5cm}
\includegraphics{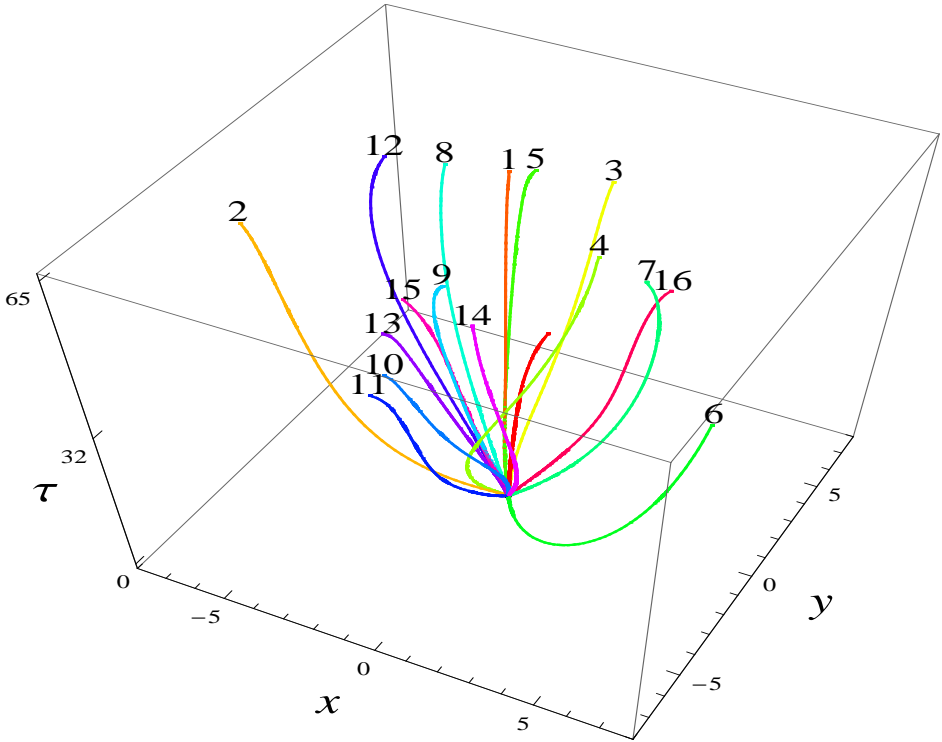}
\caption{\protect{\label{Fig-7}}Flow of the intersection points for the
  initial curves depicted in Fig.\,\ref{Fig-4}.}
\end{figure}

The maximal-size ensemble $E_{M=16}$ at $\tau=0$ is depicted 
in Fig.\,\ref{Fig-4} with the universal choice $A_0=200\,\pi$. The curves 
in Fig.\,\ref{Fig-4} are arranged in a $T$-ordered way. We have 
$T_{i=1}=65\ge T_{i=2}\ge\cdots\ge T_M=43$. 
In Fig.\,\ref{Fig-5} the evolution of an initial curve 
under curve shrinking is shown from two view points. The flow is started at $\tau=0$ and
stopped at a value of $\tau$ shortly below $T$.  
In Fig.\,\ref{Fig-7} the flow of the intersection points, corresponding
to the initial curves depicted in Fig.\,\ref{Fig-4}, is shown. 
The search for solutions to the second-order PDE Eq.\,(\ref{CSta}) subject to periodic boundary
conditions in the curve parameter,
$\hat{x}(u=0,\tau=0)=\hat{x}(u=2\pi,\tau=0)$, 
and for the initial conditions $\hat{x}(u,\tau=0)$
depicted in Fig.\,\ref{Fig-4} was performed numerically using the method of
lines. That is, the PDE was discretized on a uniform
grid in the parameter $u$ yielding a semi-discrete problem in terms of a
system of ODEs in $\tau$ which was solved using Mathematica. Fig.\,\ref{Fig-5} indicates why this technique is called
the numerical method of lines. As one can also see from Fig.\,\ref{Fig-5},
a set of discrete points on the curve, although remaining equidistant in $u$, may evolve under the
flow such that the spatial distances between next-neighbours-points falls below the
numerical precision. Numerically, the flow then encounters a singularity (not to confused
with the earlier mentioned nonfictitious singularities). To recognize such a situation
automatically, Eq.\,(\ref{diffA}) was exploited: The evolution
was stopped as soon as a sizable deviation occured from what
Eq.\,(\ref{diffA}) predicts. The configuration obtained at this point in
$\tau$ was fitted in such a way that a new discretization in $u$ yielded 
well separated points to re-start the methods of lines. 
Eq.\,(\ref{diffA}) was also used as an indicator for the final
singularity at $T$ where $A_1$ or $A_2$ or both vanish.


\subsection{Renormalization-group invariance of partition function}

\begin{figure}
\centering
\vspace{8.2cm}
\includegraphics{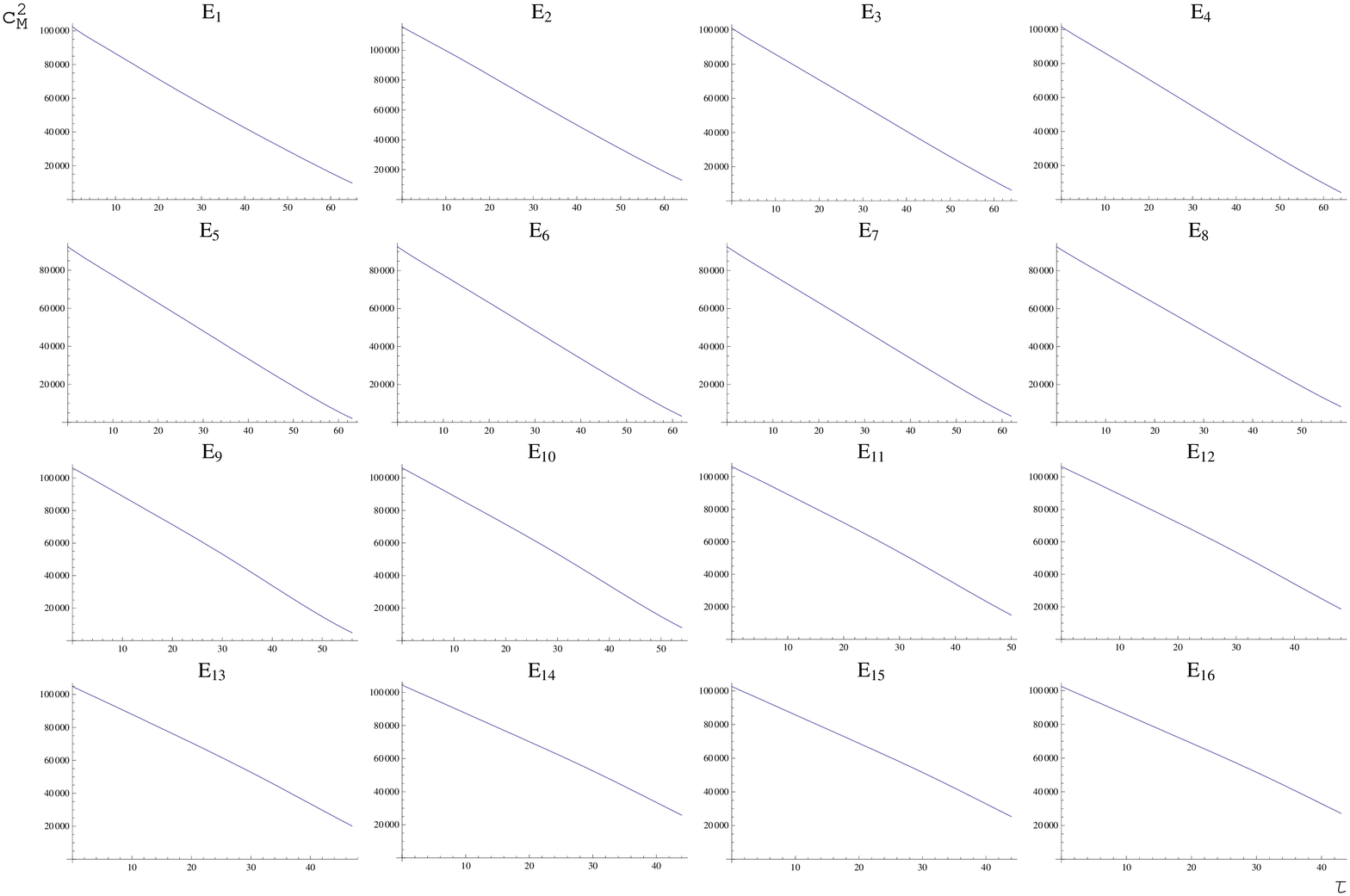}
\caption{\protect{\label{Fig-6a}}The squares of the coefficients 
  $c_M(\tau)$ entering the ansatz for effective `action' of
  Eq.\,(\ref{effectactdef}) specializing to Eq.\,(\ref{explFnc}) for
  $T$-ordered ensembles up to $M=16$.}
\end{figure}

\begin{figure}
\centering
\vspace{8cm}
\includegraphics{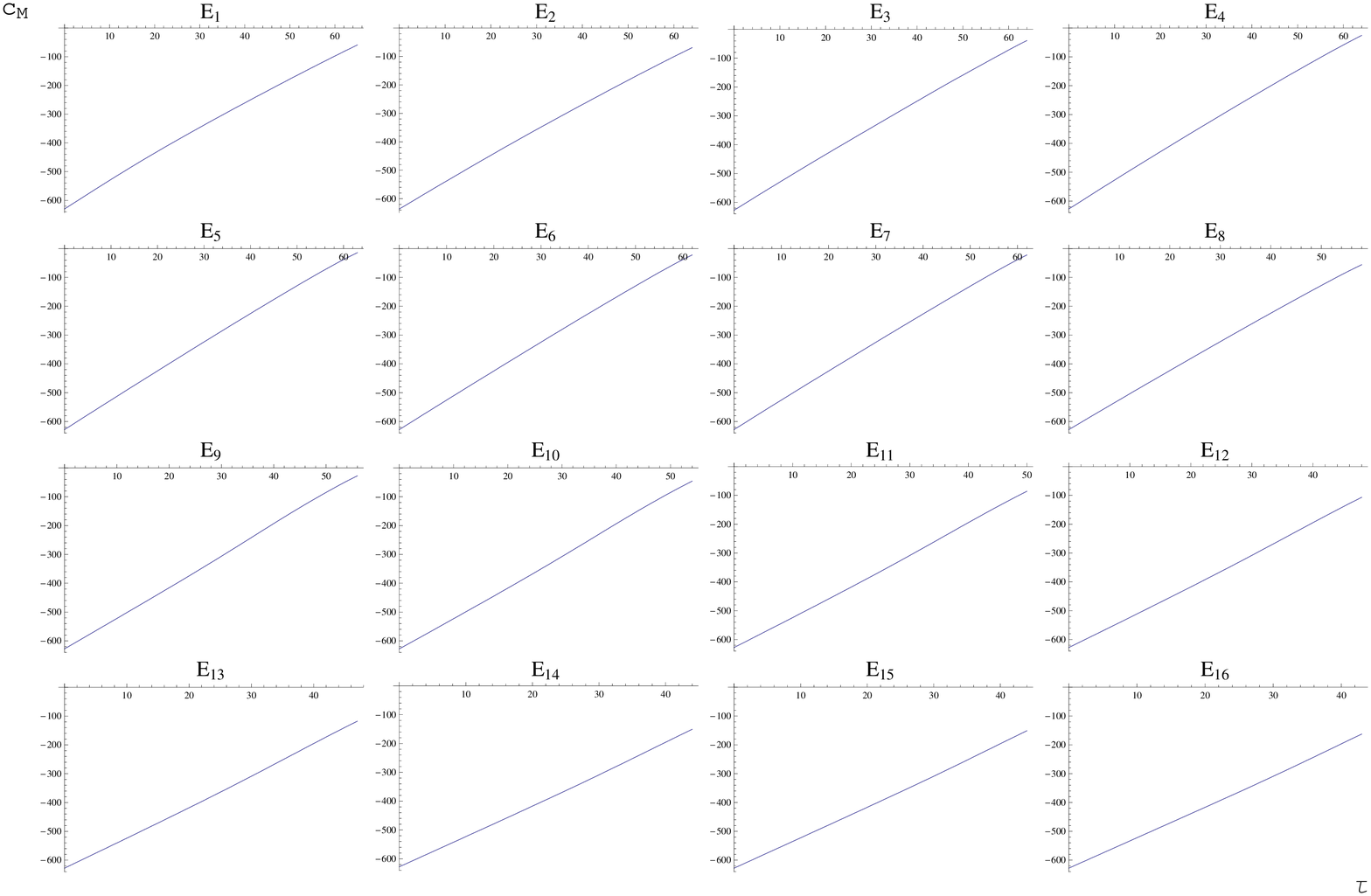}
\caption{\protect{\label{Fig-6b}}The coefficient
  $c_M(\tau)$ entering the ansatz for the effective `action' of
  Eq.\,(\ref{effectactdef}) specializing to Eq.\,(\ref{explFncmod}) for
  $T$-ordered ensembles up to $M=16$.}
\end{figure}

For all ensembles $E_M$ the $\tau$ dependence of the coefficient $c_M$
in Eq.\,(\ref{explFnc}) roughly behaves like a square root 
$\propto\sqrt{T_M-\tau}$ where  $T_M$ is the weakly ensemble-dependent minimal 
resolution. For the modified `action' 
$S_M=\frac{L(t)^2}{A(t)}\left (1+\frac{c_M(t)}{A(t)}\right)$ 
the coefficient $c_M$ is well approximated by a linear function $\propto
T_M-\tau$. Again, $T_M$ denotes a weakly ensemble-dependent minimal 
resolution. For $T$-ordered
ensembles the results for $c_M$ for the `actions' Eq.\,(\ref{explFnc}) and 
Eq.\,(\ref{explFncmod}) are shown in Figs.\,\ref{Fig-6a} 
and \ref{Fig-6b}, respectively. The results for ensembles
$E^\prime_M$ do not differ sizably from those presented in
Figs.\,(\ref{Fig-6a}), (\ref{Fig-6b}).

\subsection{Variance of location of selfintersection\label{var}}

\begin{figure}
\centering
\vspace{8.0cm}
\includegraphics{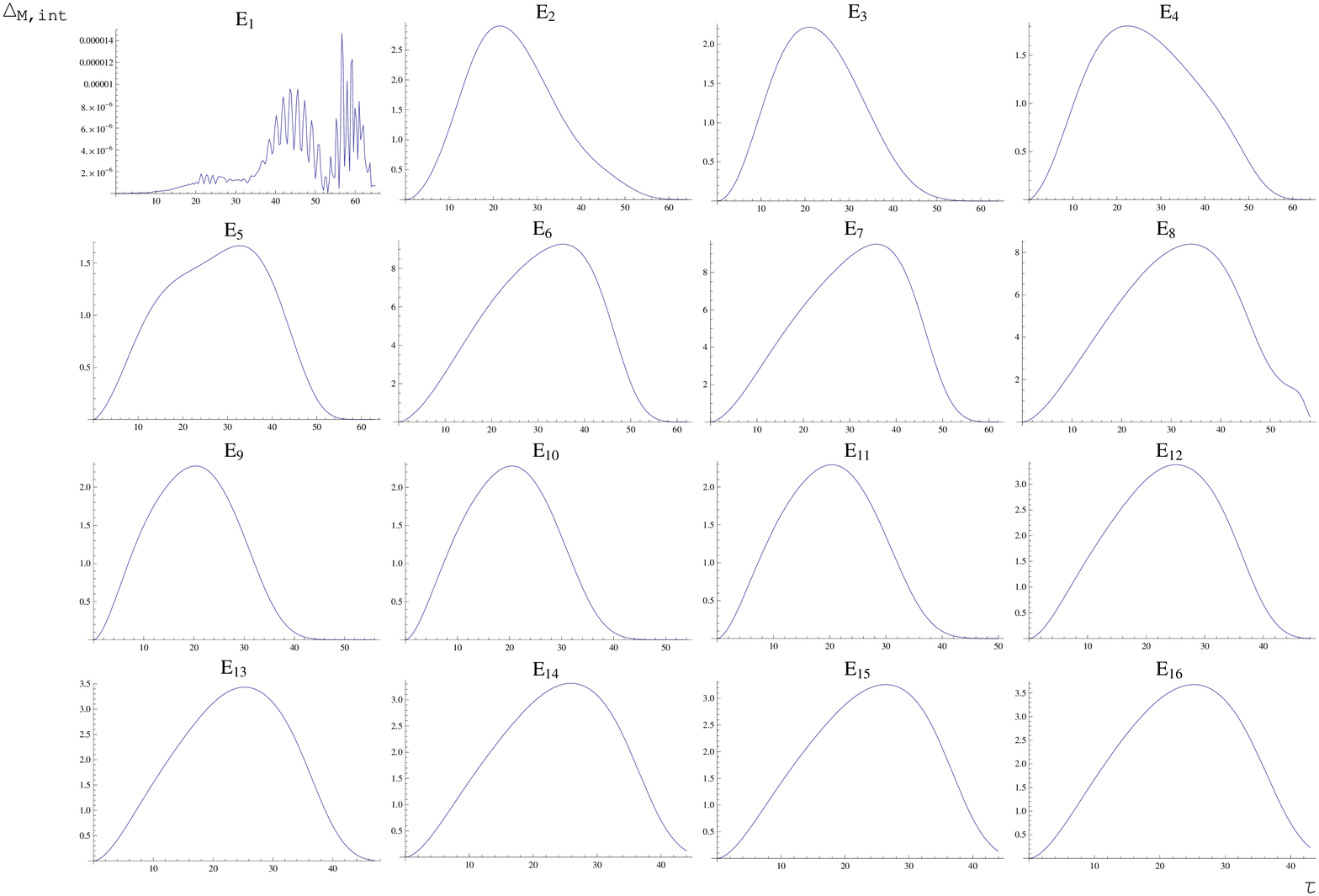}
\caption{\protect{\label{Fig-8a}}Plots of $\Delta_{M,\tiny\mbox{int}}(\tau)$ 
  for the $T$-ordered ensembles $E_M$ with $M=1,\cdots,16$. We have
  employed the ansatz for the `action'
  $S_M=\frac{L(\tau)^2}{A(\tau)}\left(1+\frac{c_M(\tau)}{L(\tau)}\right)$.}
\end{figure}

\begin{figure}
\centering
\vspace{10.4cm}
\includegraphics{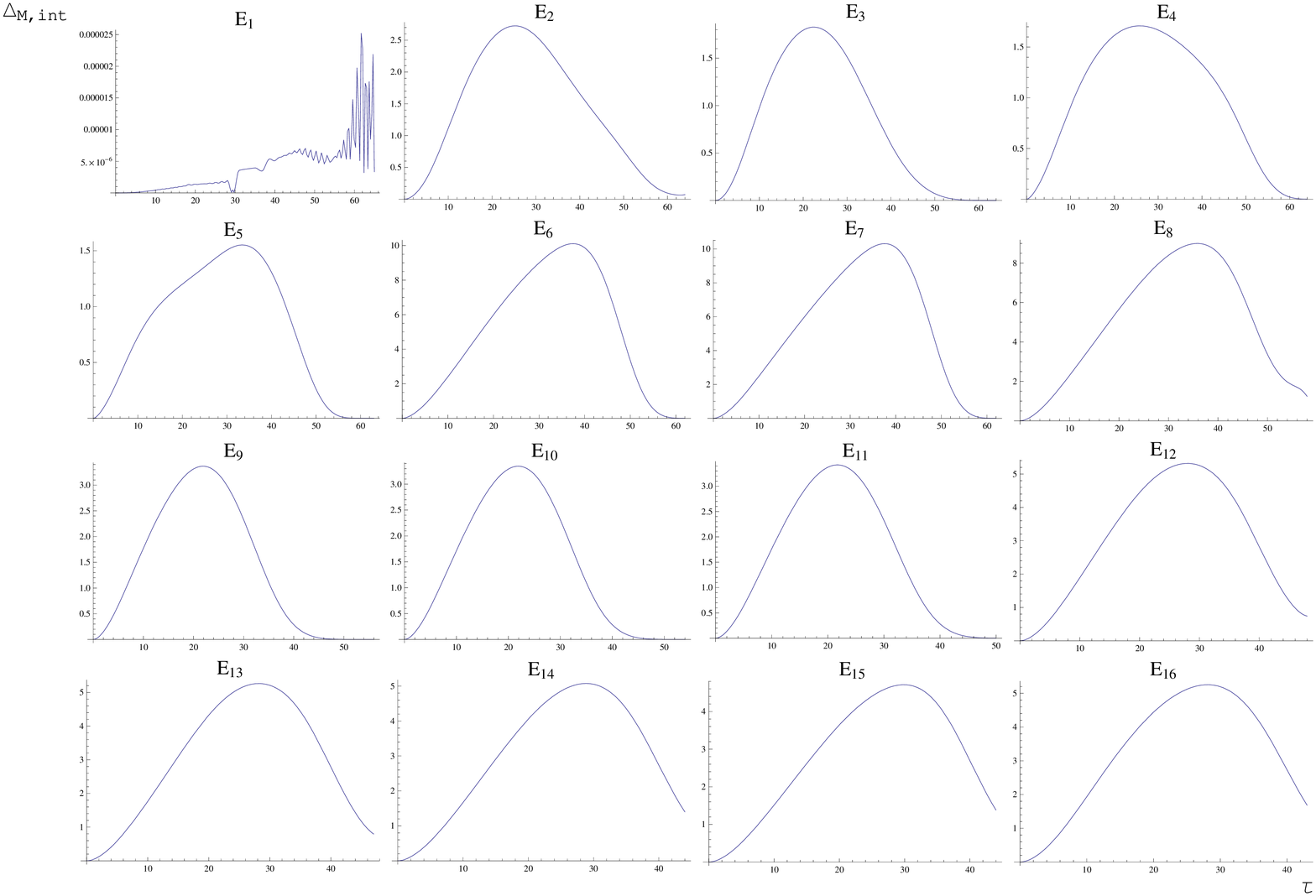}
\caption{\protect{\label{Fig-8b}}Plots of $\Delta_{M,\tiny\mbox{int}}(\tau)$ 
  for the $T$-ordered ensembles $E_M$ with $M=1,\cdots,16$. We have
  employed the ansatz for the `action'
  $S_M=\frac{L(\tau)^2}{A(\tau)}\left(1+\frac{c_M(\tau)}{A(\tau)}\right)$.}
 \end{figure}

The mean intersection $\bar{\hat{x}}_{\tiny\mbox{int}}$ over the ensemble
$E_M$ is defined as
\eqb
\label{menacom}
\bar{\hat{x}}_{\tiny\mbox{int}}(\tau)\equiv\frac{1}{Z_M}\sum_{i=1}^M
\hat{x}_{\tiny\mbox{int},i}(\tau)\exp\left(-S_M[\hat{x}_i(\tau)]\right)\,,
\eqe
where $\hat{x}_{\tiny\mbox{int},i}(\tau)$ is the location of
selfintersection (intersection point) of curve $\hat{x}_i$ at $\tau$. 
The scalar statistical deviation 
$\Delta_{M,\tiny\mbox{int}}$ of $\bar{\hat{x}}_{\tiny\mbox{int}}$ 
over the ensemble $E_M$ is defined as
\eqb
\label{statvar}
\Delta_{M,\tiny\mbox{int}}(\tau)\equiv \sqrt{\mbox{var}_{M,\tiny\mbox{int};x}(\tau)+
\mbox{var}_{M,\tiny\mbox{int};y}(\tau)}\,,
\eqe
where 
\eab
\label{vardef}
\mbox{var}_{M,\tiny\mbox{int};x}&\equiv&\frac{1}{Z_M}\sum_{i=1}^M
\left(x_{\tiny\mbox{int},i}(\tau)
-\bar{x}_{\tiny\mbox{int}}(\tau)\right)^2\,\exp\left(-S_M[\hat{x}_i(\tau)]\right)\nonumber\\ 
&=&-\bar{x}^2_{\tiny\mbox{int}}(\tau)+\frac{1}{Z_M}\sum_{i=1}^M x^2_{\tiny\mbox{int},i}(\tau)\,
\exp\left(-S_M[\hat{x}_i(\tau)]\right) 
\eae
and similarly for the coordinate $y$. In Fig.\,\ref{Fig-8a} plots of
$\Delta_{M,\tiny\mbox{int}}(\tau)$ are shown when evaluated over the ensembles
$E_1,\cdots,E_{16}$ subject to the `action' 
\dmb
S_M=\frac{L(\tau)^2}{A(\tau)}\left(1+\frac{c_M(\tau)}{L(\tau)}\right)
\dme
and the initial condition
$\bar{L}_M(\tau=0)=\tilde{L}_M(\tau=0)$. In Fig.\,\ref{Fig-8b} the
according plots of $\Delta_{M,\tiny\mbox{int}}(\tau)$ are depicted as 
obtained with the `action'
\dmb 
S_M=\frac{L(\tau)^2}{A(\tau)}\left(1+\frac{c_M(\tau)}{A(\tau)}\right)
\dme
and subject to the initial condition
$\bar{L}_M(\tau=0)=\tilde{L}_M(\tau=0)$.
Relaxing the constraint of $T$-ordering ($E_M\to E^\prime_M$) does 
not entail a qualitative change of the results.  

The results presented in Figs. \,\ref{Fig-8a}, \ref{Fig-8b} are
unexpected since in the $N=0$ sector the variance of the
'center-of-mass' saturates rapidly to finite
values. In contrast, for the $N=1$ sector the variance of the location of
selfintersection initially increases, reaches a maximum, and decreases
to zero at a {\sl finite} value of $\tau$. This is readily confirmed 
by the evaluation of the entropy, see Sec.\,\ref{entr}.  

\subsection{Evolution of entropy\label{entr}}

\begin{figure}
\centering
\vspace{8cm}
\includegraphics{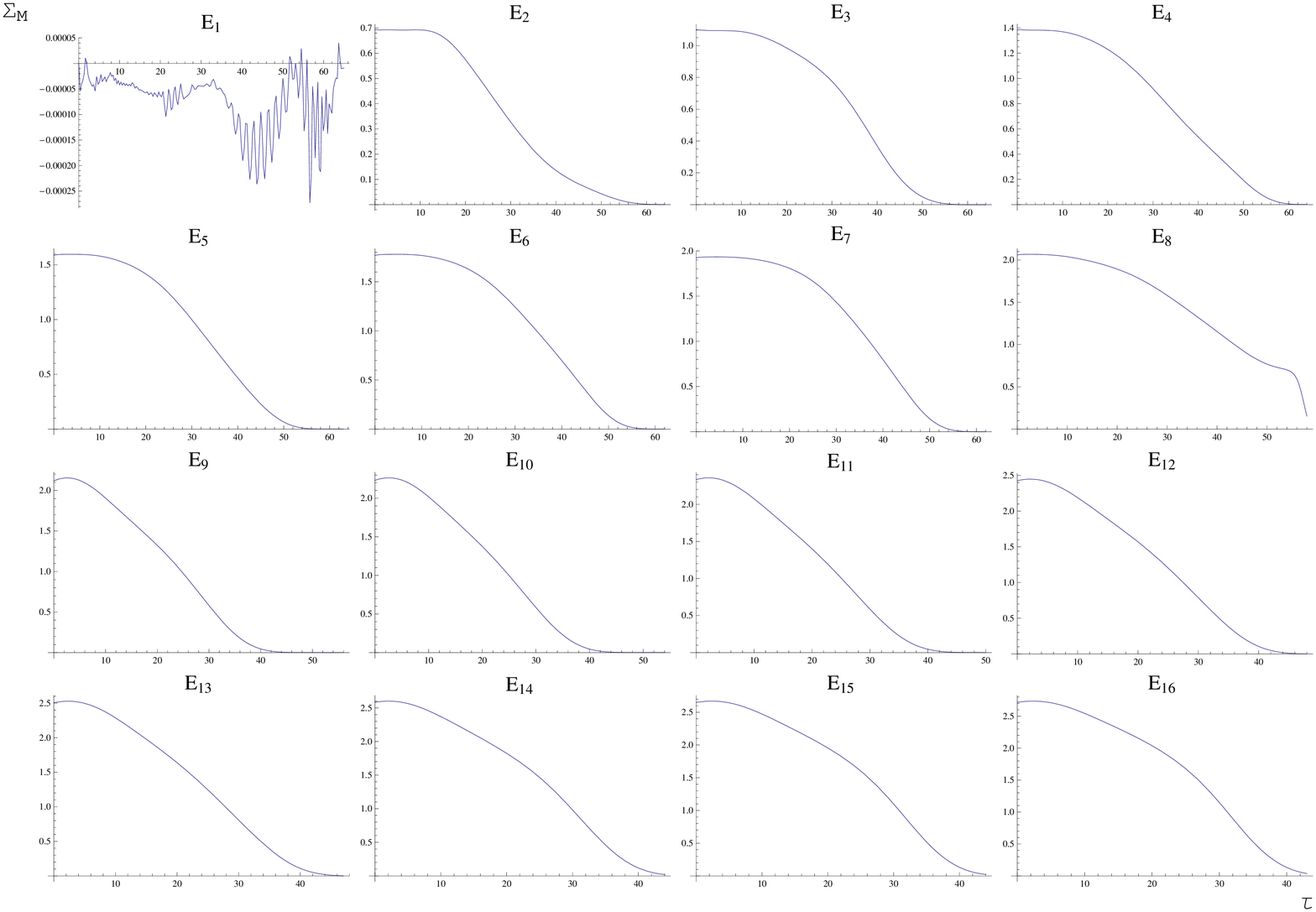}
\caption{\protect{\label{Fig-9}} Flow of the entropies $\Sigma_M$ for 
$T$-ordered ensembles of size $M=1,\cdots,16$ when evaluated with the `action'
  $S_M=\frac{L(\tau)^2}{A(\tau)}\left(1+\frac{c_M(\tau)}{L(\tau)}\right)$. The situation does not change 
qualitatively if the `action'
$S_M=\frac{L(\tau)^2}{A(\tau)}\left(1+\frac{c_M(\tau)}{A(\tau)}\right)$
is used.}
\end{figure}

Let us now evaluate the flow of entropy $\Sigma_M$ defined as
\eqb
\label{entropydef}
\Sigma_M(\tau)\equiv\log Z_M+\frac{1}{Z_M}\sum_{i=1}^{M} \exp(-S_M[\hat{x}_i(\tau)])\,S_M[\hat{x}_i(\tau)]
\eqe
where $S_M[\hat{x}_i(\tau)]$ is given by Eq.\,(\ref{effectactdef}). 
In Fig.\,\ref{Fig-9} plots are shown for $\Sigma_M(\tau)$
$(M=1,\cdots,16$ when evaluated with the `action'  
$S_M=\frac{L(\tau)^2}{A(\tau)}\left(1+\frac{c_M(\tau)}{L(\tau)}\right)$
for $T$-ordered ensembles of size $M=1,\cdots,16$.
These graphs look very much alike to the ones generated 
using the `action'
$S_M=\frac{L(\tau)^2}{A(\tau)}\left(1+\frac{c_M(\tau)}{A(\tau)}\right)$.
Notice the continuous approach to zero at finite 
values of $\tau$. This implies that order emerges spontaneously in the
system with decreasing resolution: Starting at a finite value of $\tau$, 
a particular member of $E_M$ is singled out by its weight approaching 
unity. Judging from our results for the $N=0$ sector \cite{MH2008}, 
this behavior is highly unexpected. Therefore the nontrivial topology of 
$N=1$ induces qualitative differences into the coarse-graining
process.      

\section{Summary, interpretation of results and, conclusion\label{SIC}}

In this article we have investigated the spatial 
coarse-graining of CVLs, immersed in a flat 2D plane, of an SU(2) Yang-Mills 
theory being in its confining phase. The focus was on the sector with
one topologically stabilized selfintersection (existence of an isolated
magnetic charge at its location, $N=1$). We have analysed this coarse-graining
process in terms of curve ensembles generated by evolving an initial situation
under the curve-shrinking flow \cite{GageHamilton,Grayson,GraysonII}. 
The idea here is to suppose that curve shrinking 
in the parameter $\tau$ represents an exact coarse-graining of a given
initial state and to re-construct the associated ensemble-weight of the
statistical approach (exponential of effective `action') by demanding
invariance of the corresponding partition function under the flow in
$\tau$ (renormalization-group evolution). Notice that $\tau$ is related to a physical resolution, applied to
probing the system, in a strictly monotonic decreasing manner. This
resolution may be associated with a local momentum transfer exerted by
an observer or a globally defined temperature inherent to an
environment. The functional dependence of $\tau$ on these physical 
parameters depends on the given experimental situation. 
It is, however, reasonable to assume that {\sl finite} values of $\tau$
universally correspond to {\sl finite} values of these physical parameters.  

In Secs.\,\ref{var},\,\ref{entr} we have obtained the unexpected result 
that a statistical ensemble of renormalization-group evolved curves
spontaneously orders itself in the sense that, starting from finite values of $\tau$, only a
particular member of the ensemble survives the process of 2D spatial
coarse-graining. That is, the entropy attributed to the ensemble is
practically zero for sufficiently large values of $\tau$. For the
location of selfintersection (charge of an electron) this means 
that no dissipation of energy, provided by the environment, can be mediated by the monopole
situated within the core of the intersection if the 
resolution falls below a critical, {\sl finite} value. 
This result must drastically depend on the two-dimensionality of 
space and the fact that we consider the sector with $N=1$, compare with
\cite{MH2008}.  

The recently discovered, unconventional $FeAs$ systems do not appear 
to exhibit an explicit, strong correlation between the electrons contained in
their theoretically suggested, 2D-superconducting layers, see \cite{KlausBuchner2008}
and references therein. If the two-dimensional behavior of
noninteracting electrons, subject to an environment 
represented by the parameter $\tau$, indeed is described by the
coarse-graining process investigated in the present work 
then the sudden decrease of entropy that we observe at a {\sl finite}
value of $\tau$ should ultimately be connected to this particular kind of high-$T_c$ 
superconductivity. Here $\tau$ is a monotonically
decreasing function of temperature.

\section*{Acknowledgments}
We would like to thank Francesco Giacosa and Markus Schwarz for useful conversations.


\begin{thebibliography}{10}

\bibitem{Kelvin}
W. T. Kelvin and P. G. Tait, {\sl Treatise on Natural Philosophy}, 2
vols., Cambridge University Press, 1867.  

\bibitem{Muller1986} 
J. G. Bednorz and K. A. M\"uller, Z. Phys. B{\bf 64}, 189 (1986). 

\bibitem{Anderson2005}
P. W. Anderson, arXiv:cond-mat/0510053v2.\\ 
P. W. Anderson, Physica C{\bf 460-462}, 3 (2007). 

\bibitem{Zpinch}
F. Giacosa, R. Hofmann, and M. Schwarz, Mod. Phys. Lett. A{\bf 21}, 2709
(2006).

\bibitem{FaddeevNiemi}
L. Faddeev and A. J. Niemi, Nature {\bf 387}, 58 (1997).\\ 
L. Faddeev and A. J. Niemi, Phys. Rev. Lett. {\bf 82}, 1624 (1999).\\ 
L. Faddeev and A. J. Niemi, Phys. Lett. B {\bf 525}, 195 (2002).\\ 
L. Faddeev and A. J. Niemi, Nucl. Phys. B {\bf 776}, 38 (2007).

\bibitem{Hofmann2005}
R. Hofmann, Int. J. Mod. Phys. A{\bf 20}, 4123 (2005);
Erratum-ibid.A{\bf 21}, 6515 (2006).

\bibitem{MH2008}
J. Moosmann and R. Hofmann, hep-th/0804.3527 

\bibitem{GageHamilton}
M. Gage and R. S. Hamilton, J. Differential Geometry {\bf 23}, 69
(1986). 

\bibitem{Grayson}
M. A. Grayson, J. Differential Geometry {\bf 26}, 285
(1987). 

\bibitem{Hofmann2007}
R. Hofmann, arXiv:0710.0962 [hep-th].

\bibitem{GHS2006}
F. Giacosa, R. Hofmann, M. Schwarz, Mod. Phys. Lett. A{\bf 21}, 2709
(2006).\\ 
R. Hofmann, Mod. Phys. Lett. A{\bf 22} 2657 (2007).

\bibitem{GraysonII}
M. A. Grayson, Invent. math. {\bf 96}, 177 (1989). 












\bibitem{KlausBuchner2008}
H.-H. Klauss and B. B\"uchner, Physik Journal {\bf 7}, 18 (2008). 

\end{thebibliography}
\end{document}